\documentclass{PoS}

\usepackage{xspace}
\usepackage[percent]{overpic}

\newcommand{\h}{\mbox{HESS J0632+057}\xspace}

\def \hess {H.E.S.S.\xspace}

\title{Long-term gamma-ray observations of the binary HESS J0632+057 with H.E.S.S., MAGIC and VERITAS}

\ShortTitle{Observations of HESS J0632+057 with \hess, MAGIC and VERITAS}

\author{
\speaker{G.Maier}$^{a}$, 
O.Blanch$^{b}$, 
D.Hadasch$^{c}$, 
A.L\'{o}pez-Oramas$^{d}$, 
N.Komin$^{e}$,
M.Lundy$^{f}$,
D.Malyshev$^{g}$, 
J.Moepi$^{e}$, 
S.Ohm$^{a}$,  
G.P\"{u}hlhofer$^{g}$, 
R.Prado$^{a}$,
S.Schlenstedt$^{h}$, 
D.F.Torres$^{i}$, 
B.Zitzer$^{f}$ 
for the H.E.S.S.\footnote{https://www.mpi-hd.mpg.de/hfm/HESS/}, 
MAGIC\footnote{https://magic.mpp.mpg.de/}, and 
VERITAS\footnote{https://veritas.sao.arizona.edu/} \ Collaborations\footnote{for collaboration list see PoS(ICRC2019)1177} \\
\llap{$^a$} Deutsches Elektronen-Synchrotron (DESY), Zeuthen, Germany \\
\llap{$^b$} Institut de Fisica d'Altes Energies (IFAE), The Barcelona Institute of Science and Technology (BIST), Barcelona, Spain \\
\llap{$^c$} ICRR, The University of Tokyo, Tokyo, Japan \\
\llap{$^d$} Inst. de Astrof\'{i}sica de Canarias and Universidad de La Laguna, Dpto. Astrofisica, La Laguna, Tenerife, Spain \\
\llap{$^e$} School of Physics, University of Witwatersrand, Johannesburg, South Africa \\
\llap{$^f$} Physics Department, McGill University, Montreal, Canada \\
\llap{$^g$} Institut f\"{u}r Astronomie und Astrophysik, Universit\"{a}t T\"{u}bingen, T\"{u}bingen, Germany \\
\llap{$^h$} CTA Observatory gGmbH, Heidelberg, Germany \\
\llap{$^i$} ICREA \& Institute of Space Sciences (ICE, CSIC) \& Institut d'Estudis Espacials de Catalunya (IEEC), Campus UAB, Barcelona, Spain \\
E-mail: \email{gernot.maier@desy.de}}

\abstract{The gamma-ray binary HESS J0632+057 has been observed at very-high energies (E $>$ 100 GeV) for more than ten years by the major systems of imaging atmospheric Cherenkov telescopes. 
We present a summary of results obtained with  the \hess, MAGIC, and VERITAS experiments based on roughly 440 h of observations in total. 
This includes a discussion of an unusually bright TeV outburst of HESS J0632+057 in January 2018. 
The updated gamma-ray light curve now covers all phases of the orbital period with significant detections in almost all orbital phases. 
Results are discussed in context with simultaneous observations with the X-ray Telescope onboard the Neil Gehrels Swift Observatory. }

\FullConference{36th International Cosmic Ray Conference -ICRC2019-\\
		July 24th - August 1st, 2019\\
		Madison, WI, U.S.A.}

\begin{document}
 
\section{Introduction}
\h is a gamma-ray binary consisting of a  massive Be-type star (MWC 148=HD 259440) and a compact object of unknown nature (neutron star or black hole).
Gamma-ray binaries are characterised by variable emission in X-rays and gamma-rays and a peak in their spectral energy distribution at MeV-TeV energies (see \cite{Dubus:2013} for an extensive review).
High-energy emission mechanisms and conditions under which particle acceleration in these system can take place are under debate with two major scenarios to be found in the literature:
models proposing the acceleration of particles in shocks formed by the interaction of the stellar wind with the relativistic pulsar wind ('pulsar powered', \cite{Tavani:1997}) or in shocks inside accretion-powered jets or jet-stellar window interaction zones ('accretion powered', \cite{Mirabel:1998}).
The orbital movement of the two stars results in dynamical changes of the physical conditions for particles acceleration, for the density of low-energy target photons available for gamma-ray emission through inverse-Compton scattering, and for the absorption of gamma rays by pair production.
Intensity, spectral shape, and orbital variability pattern of the observed high-energy gamma-ray emission depend therefore on a large number of parameters, among them the orbital period of the binary, the geometry of the orbit (e.g.~eccentricity and orientation), the characteristics of the stellar wind, the presence of a circumstellar disk, and the pulsar wind or jet properties.

The gamma-ray binary \h was serendipitously discovered with the \hess\ experiment during observations of the Monoceros Loop supernova remnant in 2004 and 2005 \cite{Aharonian:2007}.
It is located at a distance  of 1.1-1.7 kpc \cite{Aragona:2010}.
The orbital period of the binary was derived first from X-ray observations obtained with Swift XRT \cite{Bongiorno-2011, Aliu:2014} ($\approx$315-320 days).
Studies based on radial spectroscopy measurements at optical wavelengths provide two orbital solutions \cite{Casares:2012, Moritani:2018} for the system, which differ substantially from each other in their results for eccentricity, orientation and inclination.
The impact of this uncertainty should be taken into account in the modelling of the physical conditions at the particle-acceleration and photon-emission sites.

\h has been observed extensively in X-rays and high energies with Swift XRT \cite{Bongiorno-2011, Aliu:2014}, \hess \cite{Aharonian:2007, Aliu:2014}, MAGIC \cite{Alekscic-2016} , and VERITAS \cite{Acciari:2009, Aliu:2014}.
These observations revealed periodical flux modulations including two emission maxima and a distinct minimum visible in both the X-ray and gamma-ray observations.
\h is a very weak gamma-ray source in the MeV to GeV energy range and was only recently detected with \emph{Fermi-LAT} (Large Area Telescope) \cite{Li-2017}.
We present in these proceedings new results from observations of the gamma-ray binary HESS J0632+057 with \hess, MAGIC, and VERITAS over a period of 15 years from 2004 to 2019.
We refer to an upcoming publication for detailed discussion of the observations and interpretation of the results.


\section{Gamma-ray and X-ray Observations}

The observation of \h\ with \hess, MAGIC, VERITAS since 2003 totals to $\approx$440 hours, see Table \ref{table:ObservationSummary} for details.
The three instruments are very similar and use the technique of imaging atmospheric Cherenkov emission from extensive air showers to detect high-energy gamma rays.
The High Energy Stereoscopic System (H.E.S.S.) consists of five imaging atmospheric-Cherenkov telescopes (four 12 m-diameter telescopes and one 28 m telescope) located in the Khomas highlands of Namibia \cite{Hinton-2004}.
The MAGIC experiment is a stereoscopic system of two 17m diameter Imaging Atmospheric Cherenkov Telescopes situated on the Canary Island La Palma (Spain). 
VERITAS is an array of four imaging atmospheric-Cherenkov telescopes installed at the Fred Lawrence Whipple Observatory in southern Arizona \cite{Weekes-2002}.
For detailed descriptions of the performances of the three instruments see \cite{Alekscic-2016, Park-2017, Parson-2014}, 

Observations of \h were carried out during dark sky
and moderate moonlight conditions (moon illumination $<35$\%), with the exception of some VERITAS observations with bright moonlight conditions (see Table\ref{table:ObservationSummary}).
The latter observation mode results in a somewhat reduced sensitivity and higher energy threshold compared to observations under dark conditions.
The cadence of observations was mostly motivated to obtain a complete coverage of the long orbital light curve of the binary over several years and by technical and scheduling constrains. 

At X-ray energies, \emph{Swift-XRT} monitored HESS J0632+057 at 0.3-10 keV from 2009 January to 2019 January. The observations have typical durations of $\approx$4-5 ks taken at intervals between one week and several months. 
The data has been analysed using standard X-ray tools\footnote{The X-ray data were analysed with \texttt{heasoft v.6.22} software package and reprocessed with \texttt{xrtpipeline v.0.13.4}.}.
Fluxes were extracted from spectral fits applying an absorbed power-law model \texttt{(cflux*phabs*po)} using \texttt{XSPEC v.12.9.1m}.
Additionally, much more infrequent observations with the Chandra, XMM-Newton, and \emph{NuSTAR} observatories are included in the analysis.

All significances, fluxes, and spectral analyses are calculated using the position of the X-ray source XMMU J063259.3+054801  (Hinton et al. 2009).

%
%
%
%

\begin{center}
\begin{table}[!htb]
\begin{tabular}{lcccc}
{Observatory} &
{Observation} &
{Range of} &
{Range of } &
{Observation Time} \\
{} &
{Type} &
{Elevations} & 
{Energy Thresholds} &
{(min)} \\
{} &
{} &
{(deg)} & 
{(GeV)} &
{}  \\
\hline
\hline
VERITAS  & V4 & 56--64 & 215-325 &  1160 \\
VERITAS  & V5 &  43--64 & 200-500 &  7042 \\
VERITAS  & V6 & 44--64 & 160-400  &  6746 \\
VERITAS  & V6 red HV & 52--64 & 420-630 &  630 \\
\hline
\hess & CT1--4 & 32--62 &260-680  &  5931 \\
\hess & CT1--5 & 52--62 &200-316  &  887 \\
\hess & CT5 & 32--62 & 60-420 &  1066 \\
\hline 
MAGIC & Stereo &  38 -- 67  & 147--251  &  2880 \\
\end{tabular}
\caption{
Summary of VHE observations of HESS J0632+057.
The column observation type describes different observation modes or periods: the VERITAS observation types refer to the period before the relocation of telescope 1 as V4, the period after the relocation of telescope 1 and before the camera upgrade as V5 and after the camera upgrade as V6. Observations under bright moonlight conditions are labeled with V6 red HV.
The \hess observation types refer to the array configuration, operating either in stereo (with CT1--4 or CT1--5 cameras) or mono (observations performed with CT5 only) mode.
Poor-quality data has been removed before the calculation of the dead-time corrected observation time.
The energy threshold is defined as the lowest energy for events to be used in the analysis.
Note that this definition varies between the observatories, as different spectral reconstruction methods are applied.
\label{table:ObservationSummary}}
\end{table}
\end{center}

\section{Results}


\begin{figure}
\begin{overpic}[width=.99\textwidth]{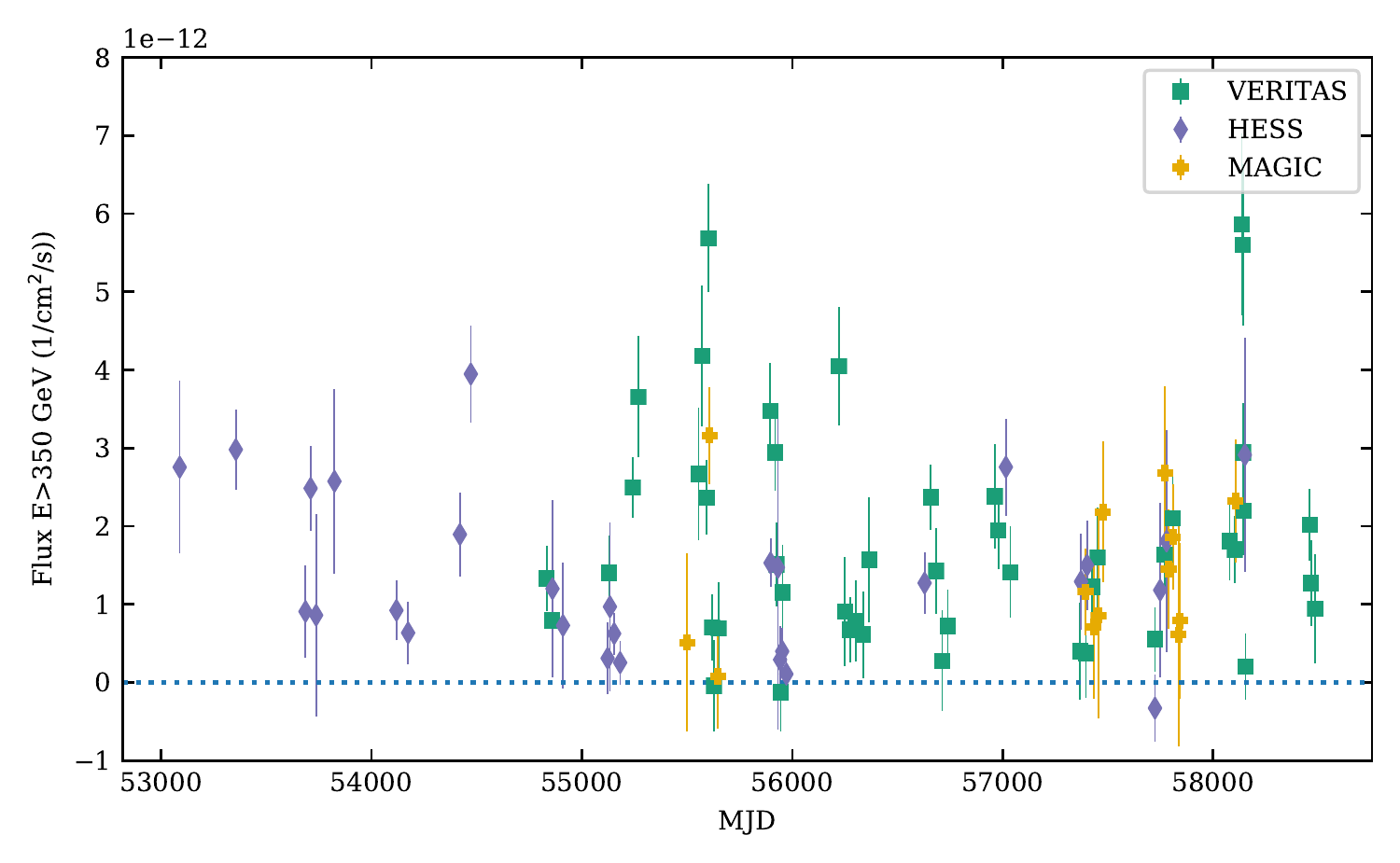}
\put (20,50) {\huge Preliminary}
\label{gammaRayLC}
\end{overpic}
\caption{Light curve of HESS J0632+057 in gamma rays (>350 GeV) for observations between 2004 and 2019 obtained with \hess, MAGIC, and VERITAS.
Each point indicates the gamma-ray flux obtained from several observations, with a time difference between individual observations of typically less than 5\% of the orbital period of 317 days. 
Vertical lines indicate one sigma statistical uncertainties.}
\end{figure}


The light curve of HESS J0632+057 in gamma rays (>350 GeV) for observations between 2004 and 2019 obtained with \hess, MAGIC, and VERITAS is shown in Figure \ref{gammaRayLC}.
The emission is variable, with  periods of flux levels below the detection limit of the instruments alternating with high-flux states reaching flux levels of $\approx$ 6\% of the flux of the Crab Nebula for energies above 350 GeV.

\begin{figure}
 \centering\begin{overpic}[width=.45\textwidth,tics=10]{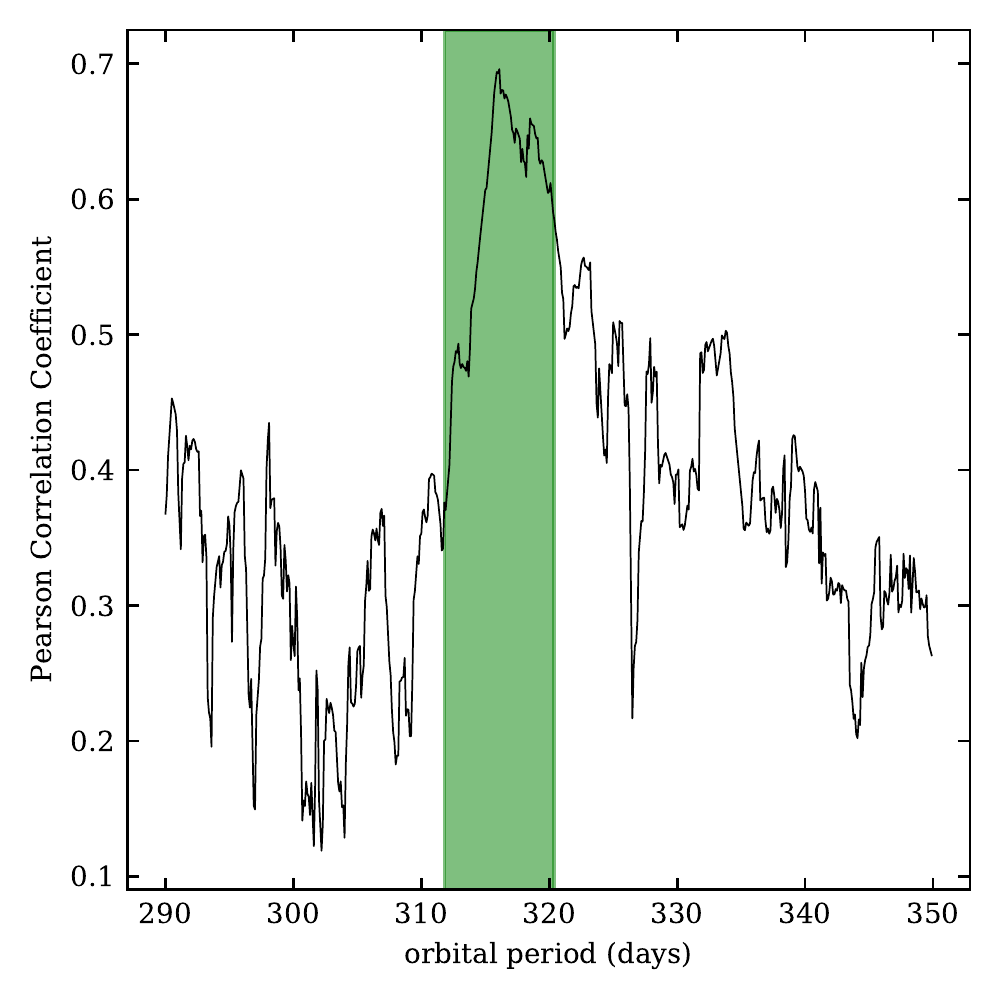}\put (60,20) {\large Preliminary}\end{overpic}
 \centering\begin{overpic}[width=.45\textwidth,tics=10]{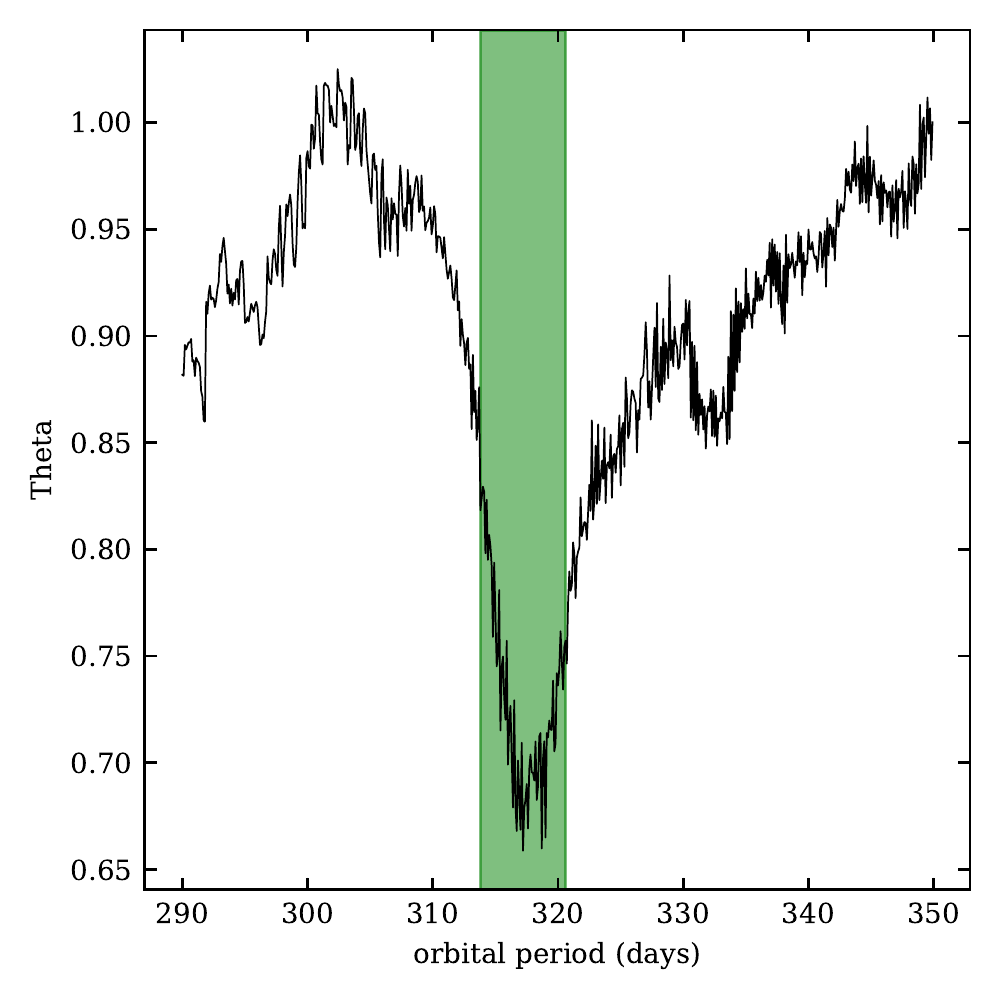}\put (60,20) {\large Preliminary}\end{overpic}
 \caption{
 Periodicity analysis of the gamma-ray light curve using the method of Pearson's correlation coefficient (left) and the phase dispersion method (right).
 Coefficients are plotted as function of assumed orbital period.
 The colored areas indicate the 68\% fiducial interval around the best estimation for the orbital period obtained by the analysis of MC-generated light curves.
 }
\label{gammaPeriod}
\end{figure}

The large data sets allows for the first time to derive the orbital period of \h from gamma-ray data. 
Figure \ref{gammaPeriod} shows the results of the periodicity analysis of gamma-ray measurements using the phase dispersion methods (PDM, \cite{Stellingwerf:1978}) and of a procedure based on the Pearson correlation coefficient (PCC, \cite{Malyshev:2017}).
The orbital period from gamma-ray data is determined to be $318.7\pm3.4$ days (PDM method) and $316.3\pm4.3$ days (PCC method), consistent with the measurements of $317.3\pm0.7$ days at X-ray energies.
Uncertainties on the orbital periods are derived from 10,000 Monte Carlo-generated light curves based on the phase-binned average profiles of the gamma-ray data.


\begin{figure}
 \centering\begin{overpic}[width=.45\textwidth,tics=10]{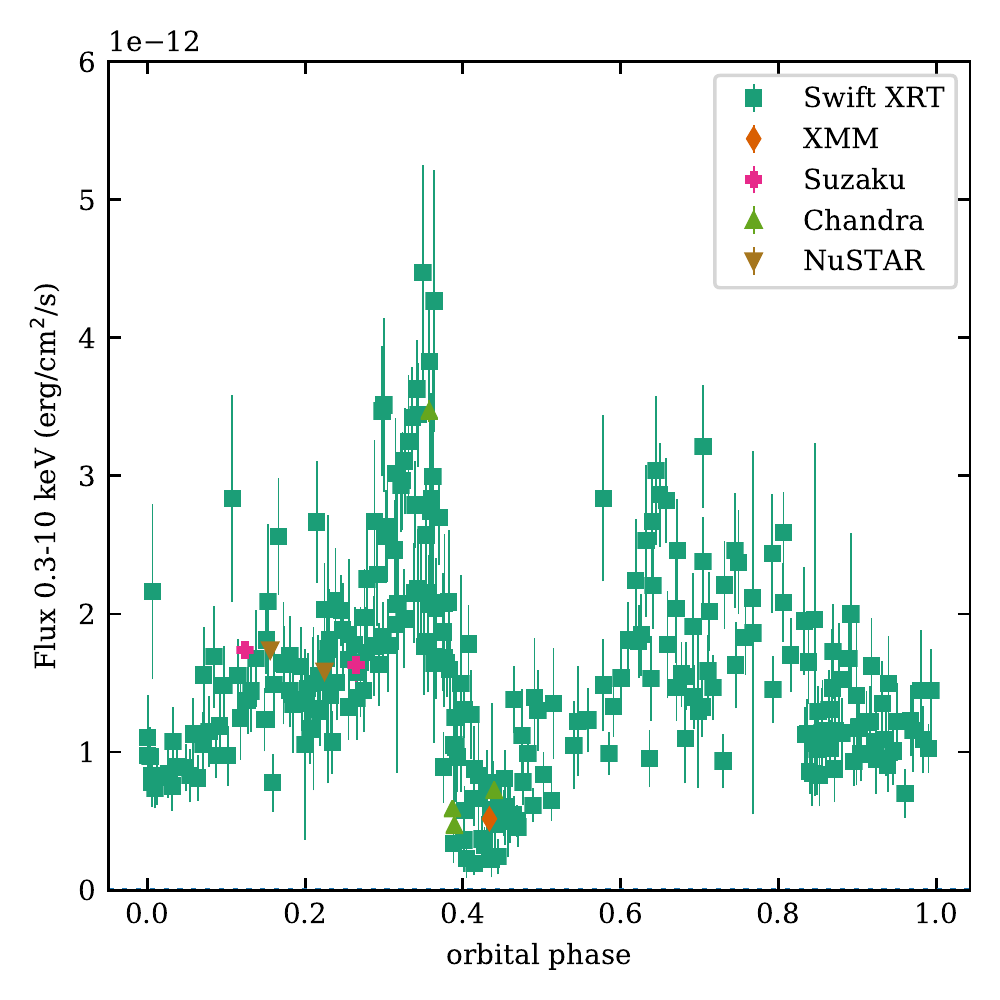}\put (60,62) {\large Preliminary}\end{overpic}
 \centering\begin{overpic}[width=.45\textwidth,tics=10]{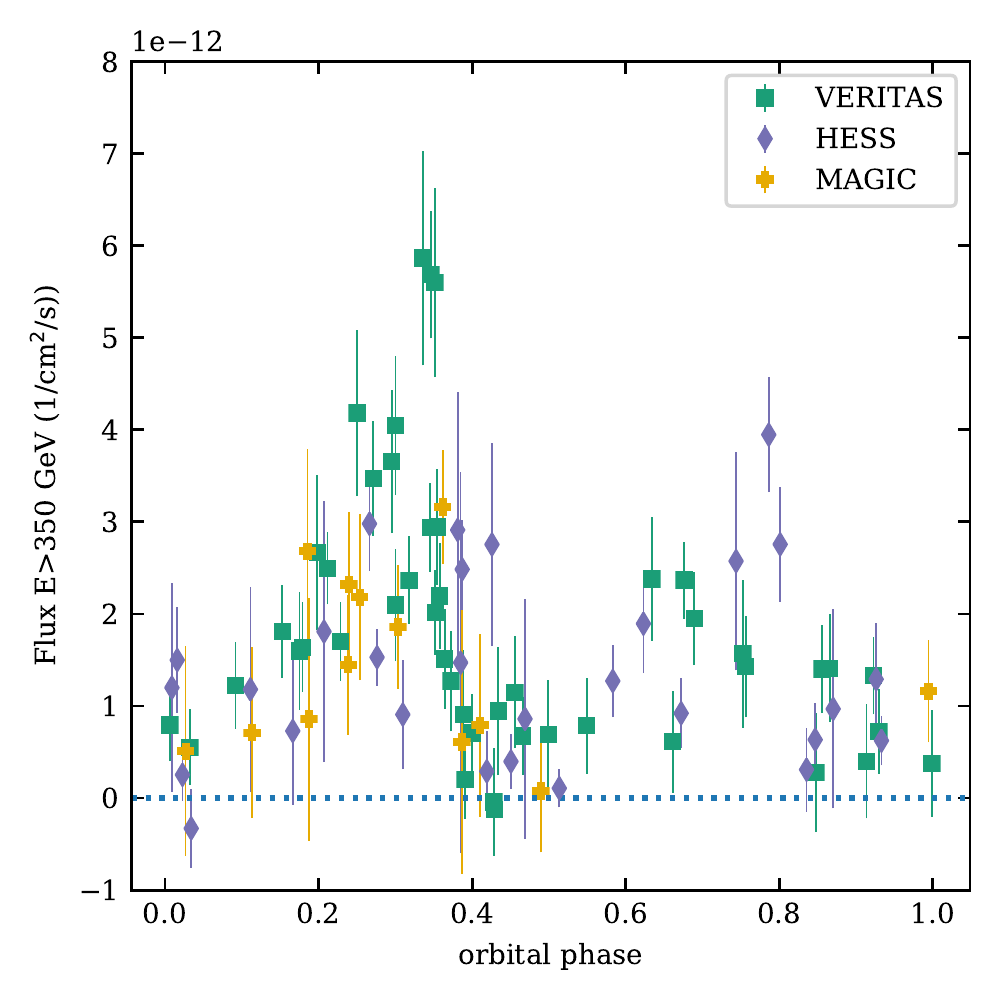}\put (60,62) {\large Preliminary}\end{overpic}
 \caption{Phase-folded X-ray (0.3--10 keV; left) and gamma-rays ($>$350 GeV; right) light curves  assuming an orbital period of 317.3 days and MJD0=54857.0 \cite{Bongiorno-2011}). 
One sigma statistical uncertainties are indicated by vertical lines (these are smaller than the marker size for all X-ray instruments but \emph{Swift-XRT}).}
\label{LCphaseFolded}
\end{figure}

The phase-folded\footnote{An orbital period of 317.3 days is assumed throughout these proceedings (derived from the available XRT observations applying an analysis similar to \cite{Aliu:2014, Malyshev:2017}).} X-ray (0.3--10 keV) and gamma-ray (energies $>$ 350 GeV) light curves  (Figure \ref{LCphaseFolded}) show very similar variability pattern across the orbit.
The correlation between X-ray and gamma-ray emission, already reported in \cite{Aliu:2014}, points towards a common origin of the radiation.
Commonly, the assumption is that X-rays are produced through synchrotron emission of high-energy particles which also produce the gamma-ray emission through inverse Compton scattering off the massive star's photon field (e.g.~\cite{Hinton-2004}).
The phase-folded gamma-ray light curve covers now all orbital phase ranges with strong detections along the orbit with the exception of the phases around 0.4-0.5. 
The  flux minimum, possibly a complete dampening of the emission, follows a decay of the emission on time scales of 1-3 weeks.
Different mechanisms are discussed for the sharp dip in the X-ray and gamma-ray emission.
This includes a complete quenching of the pulsar wind by the stellar wind during the closest approach of the two stars (assuming the orbital solution from \cite{Moritani:2018}, the passage of the compact object through the stellar disk or a significant change in the accretion rates.

\begin{figure}
\centering
\begin{overpic}[width=.49\textwidth,tics=10]{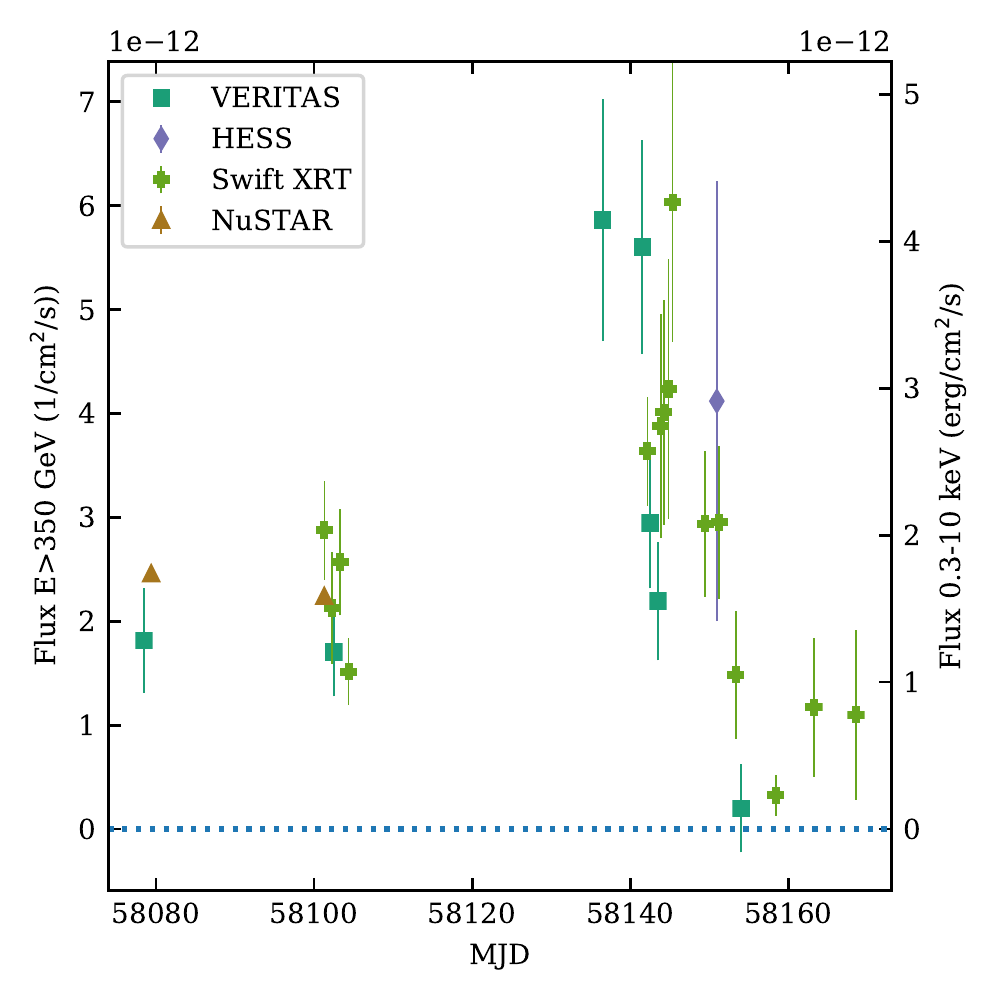}\put (20,62) {\large Preliminary}\end{overpic}
 \caption{
X-ray (0.3-10 keV; right axis) and gamma-ray (>350 GeV; left axis) light curve for the time range from 2017, Nov to 2018, March.}
\label{LCJan2018}
\end{figure}


%
%

Observations with VERITAS, \hess,  and \emph{Swift XRT} in 2018, January revealed the highest ever observed flux from this object combined with a strong decline of the flux on time scales of a few days \cite{Mukherjee:2018}.
The gamma-ray and X-ray light curve for this period is shown in Figure \ref{LCJan2018}.
Observations made with VERITAS on MJDs 58136 and 58141 detect the source with a statistical significance in excess of 10 standard deviations in a comparably short exposure of 4.3 hours only.
The gamma-ray flux is $(5.9 \pm 0.8)\times 10^{-13}$ photons cm$^{-2}$ s$^{-1}$ above 350 GeV (corresponding to about 6\% of the flux of the Crab Nebula above the same energy), about twice the flux typically measured in this orbital phase range.


\section{Conclusions}

New and updated observations with \hess, MAGIC, VERITAS and \emph{Swift XRT} of the gamma-ray binary \h  provide significantly better measurements of the complex and time-variable conditions in this gamma-ray binary. 
The data allows significantly more detailed modelling of the acceleration and emission processes.
This includes also the time-dependent contemporaneous spectral energy distributions at X- and gamma-ray energies, which will be discussed in an upcoming publication. 

\section*{Acknowledgments}

The support of the Namibian authorities and of the University of Namibia in facilitating the construction and operation of H.E.S.S.\ is gratefully acknowledged, as is the support by the German Ministry for Education and Research (BMBF), the Max Planck Society, the German Research Foundation (DFG), the Helmholtz Association, the Alexander von Humboldt Foundation, the French Ministry of Higher Education, Research and Innovation, the Centre National de la Recherche Scientifique (CNRS/IN2P3 and CNRS/INSU), the Commissariat \`{a} l'\'{e}nergie atomique et aux \'{e}nergies alternatives (CEA), the U.K. Science and Technology Facilities Council (STFC), the Knut and Alice Wallenberg Foundation, the National Science Centre, Poland grant no.~2016/22/M /ST9/00382, the South African Department of Science and Technology and National Research Foundation, the University of Namibia, the National Commission on Research, Science \& Technology of Namibia (NCRST), the Austrian Federal Ministry of Education, Science and Research and the Austrian Science Fund (FWF), the Australian Research Council (ARC), the Japan Society for the Promotion of Science and by the University of Amsterdam. We appreciate the excellent work of the technical support staff in Berlin, Zeuthen, Heidelberg, Palaiseau, Paris, Saclay, T{\"u}bingen and in Namibia in the construction and operation of the equipment. This work benefited from services provided by the H.E.S.S.\ Virtual Organisation, supported by the national resource providers of the EGI Federation. 

MAGIC would like to thank the Instituto de Astrof\'{\i}sica de Canarias for the excellent working conditions at the Observatorio del Roque de los Muchachos in La Palma. The financial support of the German BMBF and MPG, the Italian INFN and INAF, the Swiss National Fund SNF, the ERDF under the Spanish MINECO (FPA2015-69818-P, FPA2012-36668, FPA2015-68378-P, FPA2015-69210-C6-2-R, FPA2015-69210-C6-4-R, FPA2015-69210-C6-6-R, AYA2015-71042-P, AYA2016-76012-C3-1-P, ESP2015-71662-C2-2-P, CSD2009-00064), and the Japanese JSPS and MEXT is gratefully acknowledged. This work was also supported by the Spanish Centro de Excelencia ``Severo Ochoa'' SEV-2012-0234 and SEV-2015-0548, and Unidad de Excelencia ``Mar\'{\i}a de Maeztu'' MDM-2014-0369, by the Croatian Science Foundation (HrZZ) Project IP-2016-06-9782 and the University of Rijeka Project 13.12.1.3.02, by the DFG Collaborative Research Centers SFB823/C4 and SFB876/C3, the Polish National Research Centre grant UMO-2016/22/M/ST9/00382 and by the Brazilian MCTIC, CNPq and FAPERJ.

This research is supported by grants from the U.S. Department of Energy Office of Science, the U.S. National Science Foundation and the Smithsonian Institution, and by NSERC in Canada. This research used resources provided by the Open Science Grid, which is supported by the National Science Foundation and the U.S. Department of Energy's Office of Science, and resources of the National Energy Research Scientific Computing Center (NERSC), a U.S. Department of Energy Office of Science User Facility operated under Contract No. DE-AC02-05CH11231. We acknowledge the excellent work of the technical support staff at the Fred Lawrence Whipple Observatory and at the collaborating institutions in the construction and operation of the instrument.

\end{document}